\documentclass[aps,pra,twocolumn]{revtex4}
\usepackage{graphicx}
\usepackage{bm}
\begin{document}

\title{Lorenz curve of a light beam: evaluating beam  quality from a mayorization perspective}

\author{Miguel A. Porras$^1$, Isabel Gonzalo$^2$, and M. Ahmir Malik$^2$}

\affiliation{$^1$ Grupo de Sistemas Complejos, ETSIM, Universidad Polit\'{e}cnica de Madrid, Rios Rosas 21, 28003 Madrid, Spain\\
$^2$ Departamento de \'{O}ptica, Facultad de Ciencias F\'{\i}sicas, Universidad Complutense, 28040 Madrid, Spain}

%

\begin{abstract}
We introduce a novel approach for the characterization of the quality of a laser beam that is not based on particular criteria for beam width definition. The Lorenz curve of a light beam is a sophisticated version of the so-called power-in-the-bucket curve, formed by the partial sums of discretized joint intensity distribution in the near and far fields sorted in decreasing order. According to majorization theory, a higher Lorenz curve implies that all measures of spreading in phase space, and, in particular, all R\'enyi (and Shannon) entropy-based measures of the beam width products in near and far fields, are unanimously smaller, providing a strong assessment of a better beam quality. Two beams whose Lorenz curves intersect can only be considered of relatively better or lower quality according to specific criteria, which can be inferred from the plot of the respective Lorenz curves.
\end{abstract}

%
%
\maketitle
%

\section{Introduction}

The quality of a light beam is usually identified with the product of its near-field and far-field widths, normalized to that of the ideal Gaussian beam of the same wavelength, which is considered to be the best quality optical beam \cite{SIEGMAN1,WRIGHT,MORIN}. The central problem with the definition of beam quality is thus related to that of the definition of width, or spot size, of a light beam. Definitions adopted for different applications include, among many others, the $1/e^2$ decay intensity points, the circle containing 86\% of the beam power, knife-edge widths containing 10\%--90\% integrated intensity (or other numbers), and the second-moment based width or variance of the transverse intensity distribution \cite{SIEGMAN2}. In the ISO standard for beam quality measurement, the variance definition was adopted, probably because of its good analytical properties \cite{PORRAS0}, and the resulting normalized beam product factor is the popular  $M^2$ quality factor \cite{SIEGMAN1,WRIGHT,MORIN,SIEGMAN2}. Criticism of this choice was not long in coming \cite{LAWRENCE}. Variance is difficult to measure and often overestimates the beam width because of its sensitivity to noise and widespread intensity at low levels. Also, according to variance, the width of the far field intensity distribution of a uniformly illuminated slit, turns out to be infinite, which is conceptually hard to admit, and the same happens with the far-field of any aperture-truncated near-field distribution \cite{PORRAS1}. In view of these difficulties, Siegman pointed out that the near-field intensity profile together with the so-called power-in-the-bucket curve, representing then encircled power as a function of radius, contains much more useful information for an authentic assessment of beam quality \cite{SIEGMAN2}.

This problem is not exclusive to the field of optical beam characterization. It emerged much earlier in quantum mechanics in relation to the product of uncertainties in position and momentum in Heisenberg's uncertainty principle \cite{HEISENBERG}. The difficulties involved with variance as a measure of uncertainty \cite{HILGEVOORD} motivated the introduction of other measures, such as the Shannon and R\'enyi entropies \cite{SHANNON,RENYI}, that led to alternative formulations of the uncertainty principle, as their entropic formulation for position-momentum and other observables \cite{BIRULA1,BIRULA2,DEUTSCH,WEHNER}. Still, the precise formulation of the uncertainty principle, and the lowest bounds for the joint uncertainties depend on the particular measure of uncertainty, which has recently led to new formulations \cite{PARTOVI,PUCHALA,FRIEDLAND} in terms of {\it majorization} theory \cite{MARSHALL}.

In this paper we translate and adapt some of these recent developments to the optical beam area to introduce an evaluation of the quality of a light beam that is not based on specific definitions of beam width or arbitrary criteria. Majorization has recently been applied to quantify the amount of diffraction caused by uniformly illuminated apertures of different shapes \cite{LUIS}. Here we apply the same technique to the joint near-field and far-field intensity distribution of light beams having arbitrary intensity profiles in order to introduce what we call the Lorenz curve of a light beam. This curve is an adaptation of the Lorenz curve originally introduced in \cite{LORENZ}, which somehow evokes Siegman's power-in-the-bucket curve, but which incorporates the features demanded by the theory of majorization \cite{MARSHALL}. We digitize the joint near-field and far-field intensity distribution, as provided, e. g., by a CCD camera, sort their values in descending order in a single vector, and plot the accumulated intensity, or partial sums, versus phase space digitized units. Unlike the curves obtained in \cite{LUIS}, the Lorenz curve depends only on the beam shape, and not on the particular digitalization and transverse beam scaling. According to majorization theory, if the Lorenz curve of a light beam is higher than that of another beam ---the first beam majorizes the second beam---, all physically valid measures of the joint spreading in near-field and far-field, i. e., area in phase space, will be smaller for the first beam than for the second beam. This provides a robust comparison of their qualities, in the sense of being independent of the way spreading is measured.

Among all possible measures of spreading, we consider the family of R\'enyi entropies, and Shannon entropy as a particular case \cite{SHANNON,RENYI}. Given their additive property for the joint intensity distribution, their exponentials are the products of near-field and far-field entropic widths. These entropic widths are found here to be useful measures of the width that weight the different levels of intensity in the beam profile in different ways. Following majorization theory \cite{MARSHALL}, we show that the majorization relation implies that all entropic near-field and far-field beam width products are unanimously smaller for the majorizing beam than for the majorized beam.

As a strong comparison, the majorization relation between two beams is not always given. When the Lorenz curves of two beams intersect, the assessment of quality depends on the specific beam width definition, which suggests that one beam can be regarded as better than the other beam or vice versa only for specific applications. We show how to interpret intersecting Lorenz curves in terms of the compared spreading or concentration of higher and lower values in their intensity profiles. From a practical point of view, we propose to plot the Lorenz curve of a light beam together with that of a Gaussian beam for reference, whose Lorenz curve does not majorize that of all other beams (it is not the best quality beam for all purposes), but it is found that it is not majorized by any other beam.

\section{The Lorenz curve of a light beam}

We adopt a practical perspective in which a collimated light beam is focused by a lens of large enough aperture. We consider a one-dimensional and paraxial configuration for simplicity and to focus on the basic ideas. If the optical field distribution in front of the lens is $\psi(x)$, the optical field distribution at the focal plane is
\begin{equation}\label{FOCUS}
\psi'(x') =\sqrt{\frac{1}{i\lambda f}}\exp\left(\frac{i\pi x^{\prime 2}}{\lambda f}\right)\int \psi(x) \exp\left(-2\pi i \frac{x}{\lambda}\frac{x'}{f}\right)dx\, .
\end{equation}
We analyze the intensity profiles using CCD cameras of pixel sizes $\Delta x$ and $\Delta x'$ at respective pixel positions $x_j$, $j=1,2,\dots ,N$, and $x'_k$, $k=1,2,\dots ,N'$, where $N$ and $N'$ are the respective number of pixels. We may use the same camera, in which case $\Delta x= \Delta x'$ and $N=N'$. It is assumed that the CCD cameras collect almost the whole beam power, and that the beam profiles are adequately sampled. This means that the pixel readouts, $I_j \propto \int_{{\rm pixel}\, j} |\psi(x)|^2 dx\simeq |\psi(x_j)|^2\Delta x$ and $I'_k \propto \int_{{\rm pixel}\, k}|\psi'(x')|^2 dx'\simeq |\psi'(x'_k)|^2\Delta x'$, approximate the intensity distributions.

In Ref. \cite{LUIS}, majorization was applied to compare the amount of diffraction caused by different hard apertures. In few words, the normalized intensities $p_k= I'_k/\sum_{m=1}^{N'}I'_m$ at the focal plane are sorted in descending order forming the $p_k^\downarrow$ distribution, and the partial sums $S'_n=\sum_{k=1}^n p^{\downarrow}_k$, $n=1,2,\dots ,N'$ are evaluated. The focal Lorenz curve is the plot of the partial sums $S'_n$ versus pixel index $n$. Unlike the power-in-the-bucket curve, the Lorenz curve is always concave, and reaches unity at $n=N'$. The distribution $p_k$ corresponding to an aperture majorizes the distribution $\tilde p_k$ of another aperture, symbolized as $p\succ \tilde p$, if $S'_n\ge \tilde S'_n$ for all $n$, what amounts to say that the Lorenz curve $(S'_n,n)$ is higher than or equal to the curve $(\tilde S'_n,n)$. For apertures of different areas and shapes, the majorization relation $p\succ \tilde p$ is shown in \cite{LUIS} to provide a meaningful assessment, compatible with standard intuition and with previous measures of diffraction based on different criteria, that the first aperture is less diffracting than the second one. The analysis in \cite{LUIS} is limited to uniformly illuminated hard apertures of different areas and shapes, and it is assumed that the analysis is always performed with the same discretization, i. e., the same CCD camera and lens focal length. Changing the pixel size or the focal length would result in a different curve for the same physical aperture, related to the first one by a horizontal scaling factor, so that two experimentalist performing the majorization relations must first agree in these details. From a more conceptual point of view, the majorization comparison in \cite{LUIS} includes the effect of the size of the aperture on diffraction, i. e., considers a larger aperture less diffracting than (majorizes) a smaller aperture of the same shape. While this is true, it is also obvious, and we would like to have a measure of the diffraction that is independent of the size of the beam and that depends only on its shape.

To this purpose, we also consider the normalized intensity $q_j=I_j/\sum_{m=1}^N I_m$ in front of the lens, construct the {\it joint} intensity distribution $r_{j,k}=q_j p_k$, whose elements are sorted in descending order in the vector $r^\downarrow_i$, $i=1,2,\dots , NN'$. We first note that under the scaling $\psi(x)\rightarrow \psi(x/a)$ changing the beam width in front of the lens, the focal field scales as $\psi'(x')\rightarrow a\psi'(ax')$. The particular elements of $r_{j,k}$ are then generally different, but the rearrangement in decreasing order in the $r_i^\downarrow$ vector washes out the differences, except for spurious discretization effects. The partial sums $S_n=\sum_{i=1}^n r^{\downarrow}_i$, $n=1,2,\dots ,NN'$ versus $n$ define also the same curve irrespective of the beam size in front of the lens. The dependence on the pixels size, focal length and wavelength is removed by replacing the arbitrary index $n$ in the horizontal axis with physical phase space units, $n \Delta x \Delta \xi$, where $\Delta\xi=\Delta x'/\lambda f$. The word "same" means here that measurements with the same beam using different lenses and CCD cameras, would yield discrete versions of an ideal Lorenz curve in the hypothetical limit of $\Delta x,\Delta x'\rightarrow 0$ and $N,N'\rightarrow \infty$. Lorenz curves have indeed been introduced also for continuous functions \cite{MARSHALL}, and hence could directly be introduced for $|\psi(x)|^2$, $|\psi'(x')|^2$ and their product, but the mathematics involved in ordering continuous functions is cumbersome and far from the simple experimental arrangement described above.

\begin{figure*}[t!]
\centering
\includegraphics[width=5.0cm]{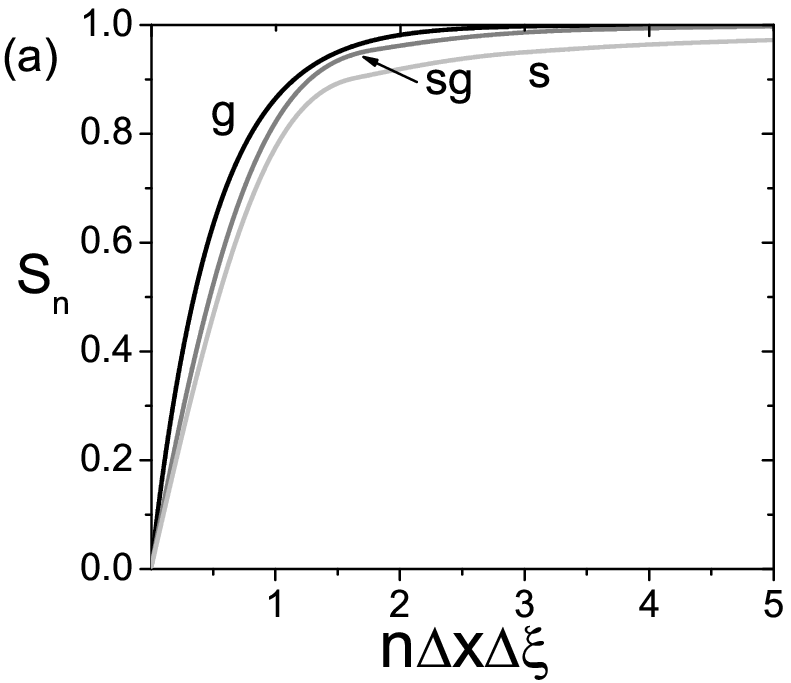}\includegraphics[width=5.0cm]{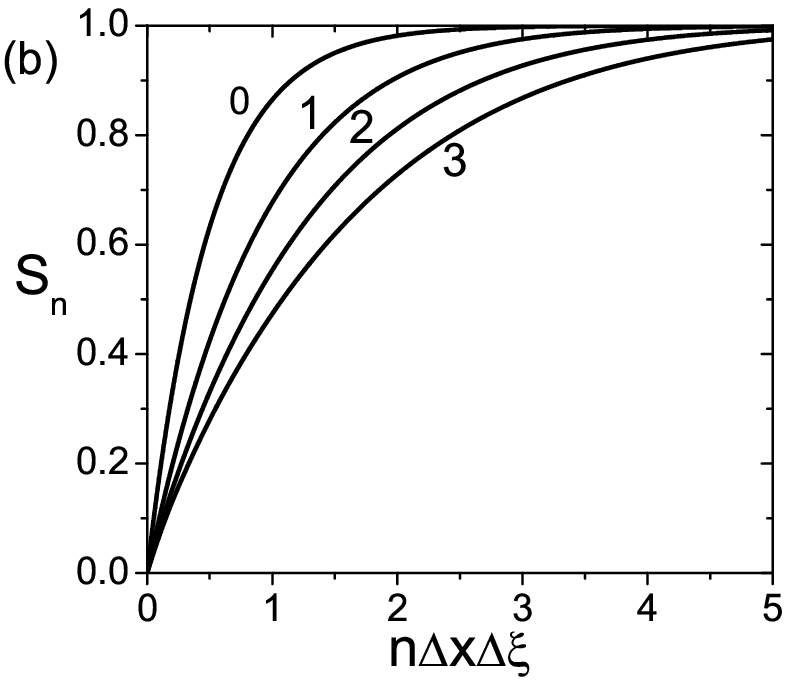}\includegraphics[width=5.0cm]{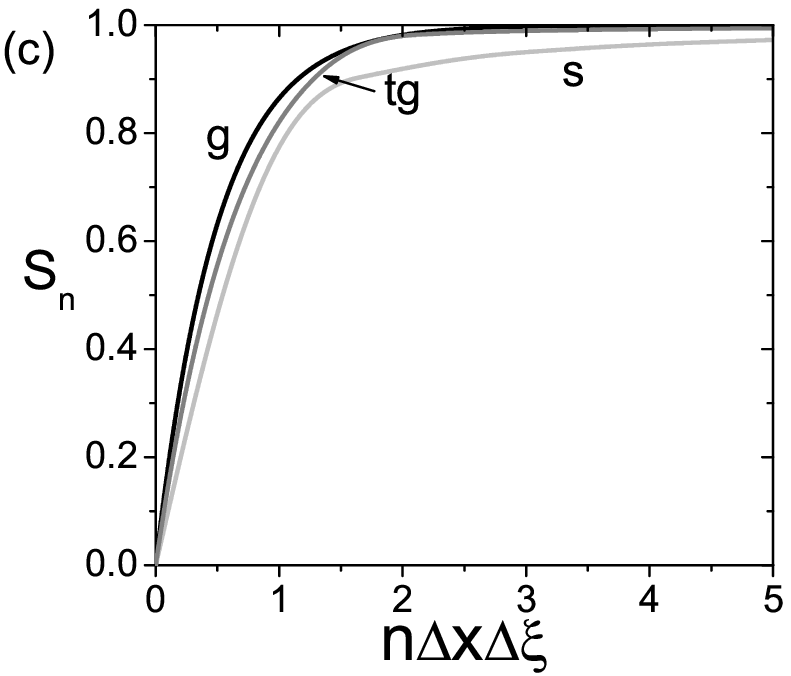}
\caption{Lorenz curves of light beams with near-field (a) Gaussian profile $\psi(x)\propto\exp(-x^2)$ [g], super-Gaussian profile $\psi(x)\propto\exp(-x^8)$ [sg], slit profile $\psi(x)\propto 1$ in $[-1,1]$ [s], with near-field (b) Hermite-Gauss profiles $\psi(x)\propto\exp(-x^2)H_m(\sqrt{2}x)$ of the indicated orders $m=0,1,2$ and $3$, and with near-field (c) truncated Gaussian profile $\psi(x)\propto\exp(-x^2)$ in $[-1,1]$ [tg] compared to those with Gaussian and of slit profiles. All curves are invariant under $x$ scaling. None of the Lorenz curves intersect.}
\label{Fig1}
\end{figure*}

As in Ref. \cite{LUIS}, but applied now to the joint intensity distribution, $r\succ \tilde r$ implies that the light beam $\psi$ is narrower than $\tilde \psi$ in phase space, i. e., in the near and far fields at the same time, and therefore can be considered of better quality. This is illustrated with a few examples in Figs. \ref{Fig1}. The Gaussian beam, usually considered the best beam, majorizes super-Gaussian beams, which in turn majorize the slit [Fig. \ref{Fig1}(a)]; Hermite-Gauss beams of increasing order are, according to majorization, increasingly worse beams as their order increase [Fig. \ref{Fig1}(b)]; the Gaussian beam majorizes truncated Gaussian beams, which in turn majorize the slit as their strong truncation limit [Fig. \ref{Fig1}(c)]. In the Appendix, we demonstrate that the above Lorenz curve reduces to that defined in Ref. \cite{LUIS} for uniformly illuminated apertures. Thus, the present Lorenz curve yields the same results regarding majorization of uniformly illuminated apertures, e. g., a number of separated slits is majorized by (diffract more than) a single slit.

We note that the above comparisons involving truncated Gausssian or different slits cannot be made on the basis of the variance-based definition of beam quality, since the second-order moments at focus diverge \cite{PORRAS1}. Instead, the Lorenz curve is well-defined for any real light beam carrying finite power ($p$ and $q$ are normalizable). Further, majorization provides a global and strong (in the sense explained below) assessment of quality that is not based on particular details of the beam profile, as the size of the main lobe, positions of first zeros, or different criteria for encircled power.

In the following we find it more convenient to use $\psi(x)$ and its Fourier transform
\begin{equation}
\phi(\xi)=\int \psi(x) \exp\left(-2\pi i x\xi \right)dx \, ,
\end{equation}
giving the equivalent discretized data
\begin{equation}\label{DIS}
q_j= \frac{|\psi(x_j)|^2\Delta x}{\sum_{m=1}^N |\psi(x_m)|^2\Delta x}\,, \quad p_k= \frac{|\phi(\xi_k)|^2\Delta \xi}{\sum_{m=1}^{N'} |\psi(\xi_m)|^2\Delta \xi} \, ,
\end{equation}
and joint distribution $r_{j,k}=q_j p_k$, and therefore the same Lorenz curve and conclusions with regard to majorization.

\section{Entropies, entropic widths and width products}

The assessment of majorization regarding uncertainty can be said to be {\it strong,} in the sense that all physically valid measures of spreading or uncertainty (the so-called Schur-concave functions) of the majorizing distribution are smaller than the respective measures of uncertainty for the majorized distribution \cite{MARSHALL}. We will see in this Section that if these measures are additive, i. e., entropies with the property that the uncertainty of $r_{j,k}$ is the sum of the uncertainties in $q_j$ and $p_k$, then there exist an infinite number of measures of width whose products in space and in spatial frequency, that is, areas in phase space, are all smaller for the majorizing distribution than for the majorized one, confirming unanimously that the first beam is of better quality than the second one.

As said, $r\succ \tilde r$ implies that for all Schur-concave functions $H$, $H(r)\le H(\tilde r)$. Among them, entropies have the additive property that $H(r)=H(q)+H(p)$. Important examples are the Shannon and R\'enyi entropies \cite{SHANNON,RENYI}. For a generic distribution $s_i$, $i=1,2,\dots , M$, the R\'enyi entropies are given by \cite{RENYI}
\begin{equation}
H_\alpha(s)= \frac{1}{1-\alpha}\ln \sum_{i=1}^M s_i^\alpha \, ,
\end{equation}
for any $\alpha\ge 0$. The particular case of the R\'enyi entropy with $\alpha=1$ is the famous Shannon entropy $H_1(s)=-\sum_{i=1}^M s_i \ln s_i$ \cite{SHANNON}. Other R\'enyi entropies of particular interest are the min-entropy $H_\infty=-\ln(\mbox{max}\, s)$ and the max-entropy $H_0=\ln M$. Any other entropy verifies
\begin{equation}\label{HIERARCHY}
H_0(s)\ge H_\alpha(s)\ge H_\infty(s)\,.
\end{equation}
In any case, for the joint distribution $r_{j,k}=q_j p_k$, one has $H_\alpha(r)=H_\alpha(q)+H_\alpha(p)$. Thus, for two light beams such that $r\succ \tilde r$,
\begin{equation}\label{ENTRO1}
H_\alpha(q)+H_\alpha(p)\le H_\alpha(\tilde q)+H_\alpha(\tilde p)\,.
\end{equation}
The R\'enyi entropies depend on the beam profiles and on the particular discretization (number and size of pixels). This can be seen using Eqs. (\ref{DIS}), writing $(\Delta x)^\alpha=(\Delta x)^{\alpha-1}\Delta x$ [an analogously for $(\Delta\xi)^\alpha$], to rewrite
\begin{equation}\label{ENTRO2}
H_\alpha(q)= - \ln \Delta x + H_\alpha(\psi) \, , \quad H_\alpha(p)= - \ln \Delta \xi + H_\alpha(\phi)
\end{equation}
where
\begin{equation}
H_\alpha(\psi) = \frac{1}{1-\alpha}\ln \sum_{j=1}^N \left(\frac{|\psi(x_j)|^2}{\sum_{m=1}^N |\psi(x_m)|^2\Delta x}\right)^\alpha \Delta x \, ,
\end{equation}
and analogously for $H_\alpha(\phi)$. Provided that the beam profile is suitably sampled and the whole power is collected, $H_\alpha(\psi)$ and $H_\alpha(\phi)$ coincide, except for small discretization errors, with the continuous R\'enyi entropies \cite{BIRULA2}
\begin{eqnarray}
H_\alpha(\psi) &=& \frac{1}{1-\alpha}\ln \int \left(\frac{|\psi(x)|^2}{P}\right)^\alpha dx \, , \\
\quad H_\alpha(\phi) &=& \frac{1}{1-\alpha}\ln \int \left(\frac{|\phi(\xi)|^2}{P}\right)^\alpha d\xi \, ,
\end{eqnarray}
where $P=\int |\psi(x)|^2 dx=\int |\phi(\xi)|^2d\xi$ is the beam power. We have then separated the entropies in Eqs. (\ref{ENTRO2}) in a term that depends on the discretization, and the continuous entropies that depend only on the beam profile. For the two light beams such that $r\succ \tilde r$,  Eq. (\ref{ENTRO1}) can then be simplified to the relation
\begin{equation}\label{ENTRO3}
H_\alpha(\psi)+ H_\alpha(\phi)\le  H_\alpha(\tilde\psi) + H_\alpha(\tilde\phi)
\end{equation}
concerning only their physical intensity profiles.

To understand the relevance of Eq. (\ref{ENTRO3}), we point out that
\begin{equation}
D_\alpha(\psi)= \exp[H_\alpha(\psi)]
\end{equation}
are, for all values of $\alpha$, quantities with the dimension of length that measure the space occupied by the function $|\psi(x)|^2$, full width or spot size, weighing its higher and lower values differently. From Eq. (\ref{ENTRO2}), the values of these {\em entropic widths} can directly be evaluated from the discretized data as $D_\alpha=\Delta x \exp[H_\alpha(q)]$, or equivalently, $D_\alpha=\Delta x \left(\sum_{j=1}^N q^\alpha\right)^{1/(1-\alpha)}$.

\begin{figure}
\centering
\includegraphics[width=8cm]{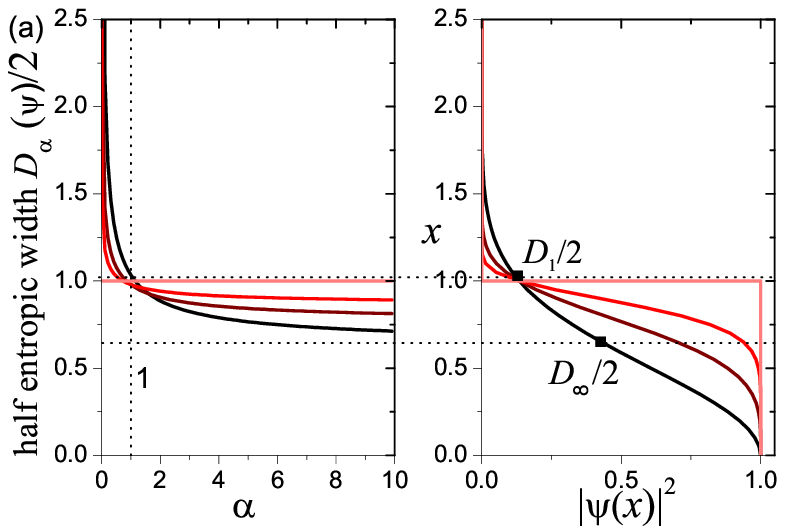}
\includegraphics[width=8cm]{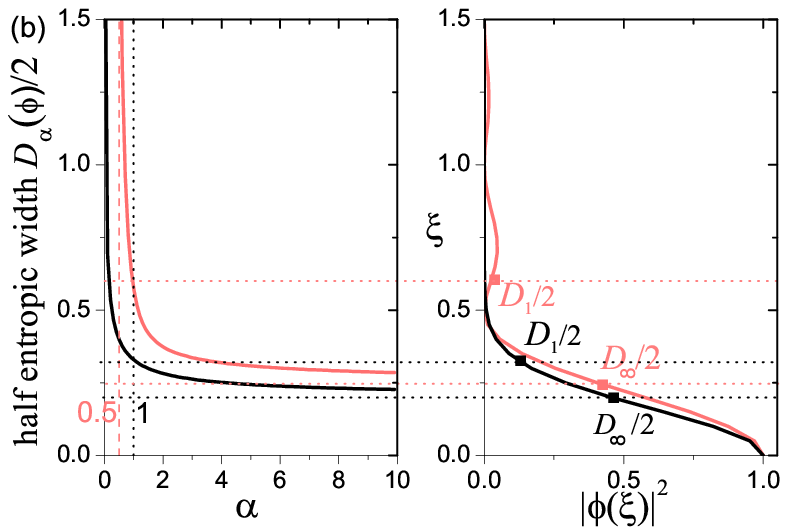}
\caption{(a) Half entropic widths, $D_\alpha(\psi)/2$, as functions of $\alpha$ (left panel) for the Gaussian profile $\psi(x)\propto \exp(-x^2/a^2)$ (black), the super-Gaussian profiles $\psi(x)\propto \exp(-x^s/a^s)$, $s=4$ and $8$ (dark red and red) and the uniform slit in $[-a,a]$ (light red), whose intensity profiles are shown in the right panel for comparison. The dots in the right panel indicate the Shannon and min-entropic width of the Gaussian. Intensities are normalized to the peak intensity for better visibility, and $a=1$ (arbitrary units) in all cases. (b) Half entropic widths, $D_\alpha(\phi)/2$, as functions of $\alpha$ (left panel) for the Fourier transforms of the above Gaussian and uniform slit, i. e., $\phi(\xi)\propto\exp[-(\pi a\xi)^2]$ (black) and $\phi(\xi)\propto \sin(2\pi a \xi)/(2\pi a\xi)$ (light red), whose intensity profiles are shown in the right panel. The dots in the right pannel indicate the Shannon and min-entropic widths of the respective profiles. Intensities are normalized to the peak intensity for better visibility, and $a=1$ as in (a).}
\label{Fig2}
\end{figure}

As a few examples, the left panels in Figs. \ref{Fig2}(a) and \ref{Fig2}(b) plot the values of the entropic widths $D_\alpha(\psi)$ and $D_\alpha(\phi)$ as functions of $\alpha$ for typical beam profiles $\psi(x)$ and for their Fourier transforms $\phi(\xi)$, plotted in the corresponding right panels. For a slit of full width $2a$ (or a set of slits of total width $2a$), all entropic widths yield $D_\alpha=2a$ [light red curves in Fig. \ref{Fig2}(a)], which seems to tell us that the width is $2a$ from any point of view. In this respect, $D_\alpha$ are clearly superior to the variance-based definition (yielding $2a\sqrt{2/3}$ for the single slit, or a quantity that depends on the separation between slits of total width $2a$, see e. g. \cite{BIRULA2}). For any other profile, and according to relation (\ref{HIERARCHY}), the entropic widths verify $D_0(\psi)\ge D_\alpha(\psi)\ge D_\infty(\psi)$. Using that $H_\infty(q)=-\ln(\mbox{max}\,q)$ and Eq. (\ref{DIS}), the min-entropic width, $D_\infty(\psi)$, is readily seen to coincide with the quotient of the power and the peak intensity, $D_\infty(\psi)=P/\mbox{max}\,|\psi|^2$. For beams with a single pronounced maximum such as the Gaussian, super-Gaussian or sinc profiles in Figs. \ref{Fig2}(a) and (b), the min-entropic width is close to the standard full-width at half-maximum, as can be seen in Fig. \ref{Fig2}. For decreasing $\alpha$, $D_\alpha$ increases because intensities lower than the peak intensity are increasingly weighted. The particular case of the Shannon entropic width ($\alpha=1$) was studied in detail in \cite{PORRAS2}, and it was found to provide adequate commitment between higher and lower intensities so that it provides values of the width close to standard measures, at the same time that it is relatively insensitive to noise and widespread intensity, and more importantly, it is well-defined for the Fourier transform of truncated beams. For example, $D_1=2.066 a$ for the Gaussian profile $\exp(-x^2/a^2)$, which is almost equal to the standard variance-based full width $2a$, or $D_1=1.144/a$ for the Fourier transform $\phi(\xi)\propto \sin(2\pi a \xi)/(2\pi a\xi)$ of the uniform beam in $[-a,a]$, which is about the distance between the two first opposite zeros, as shown in Fig. \ref{Fig2}. The entropic widths are indeed well-defined for the Fourier transform of truncated beams for $\alpha>0.5$. In the limit $\alpha\rightarrow 0$, using that $H_0(q)=\ln N$, the max-entropic width, $D_0(\psi)$, is seen to measure the support of $\psi$, meaning that all intensities are equally weighted in this limit, and that $D_0(\psi)$ is finite only for truncated profiles.

\begin{figure*}[t!]
\centering
\includegraphics[width=5.0cm]{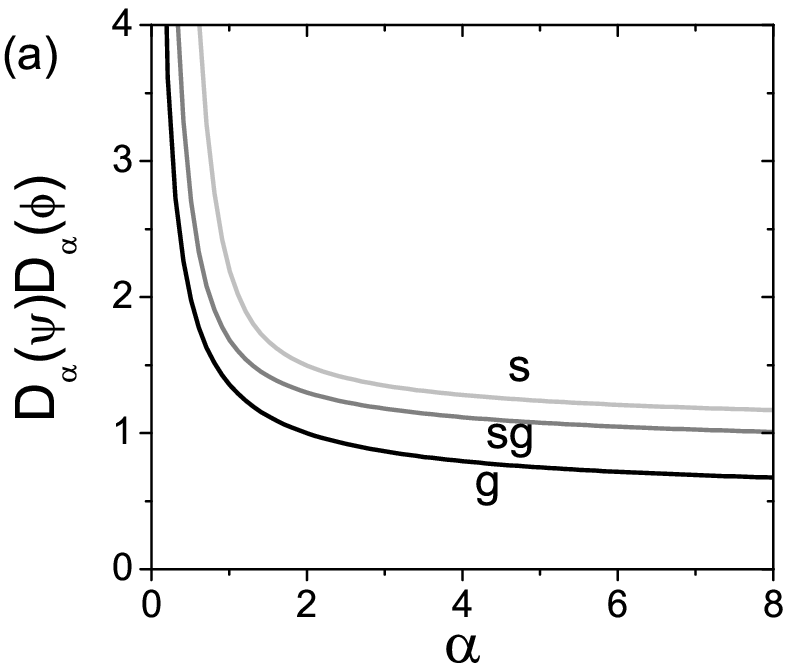}\includegraphics[width=5.0cm]{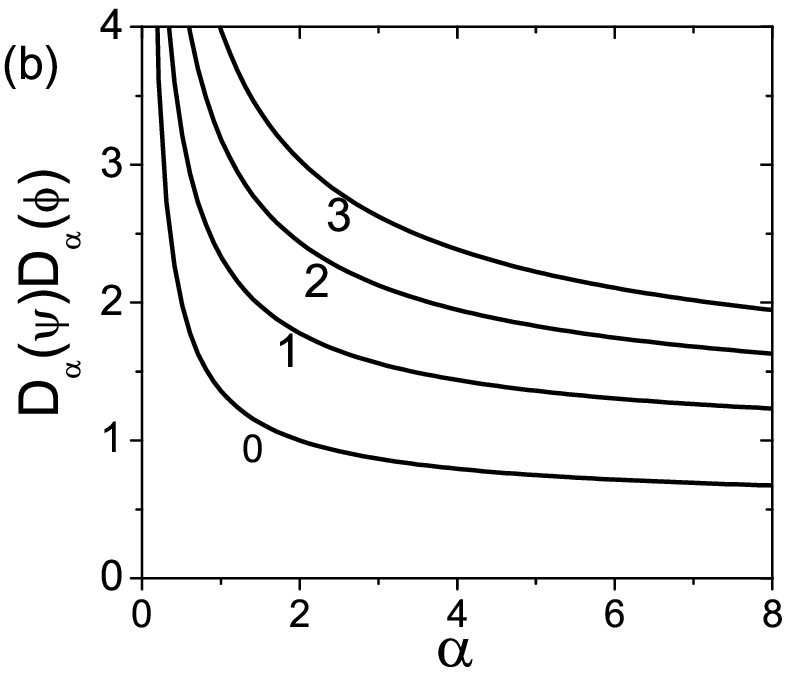}\includegraphics[width=5.0cm]{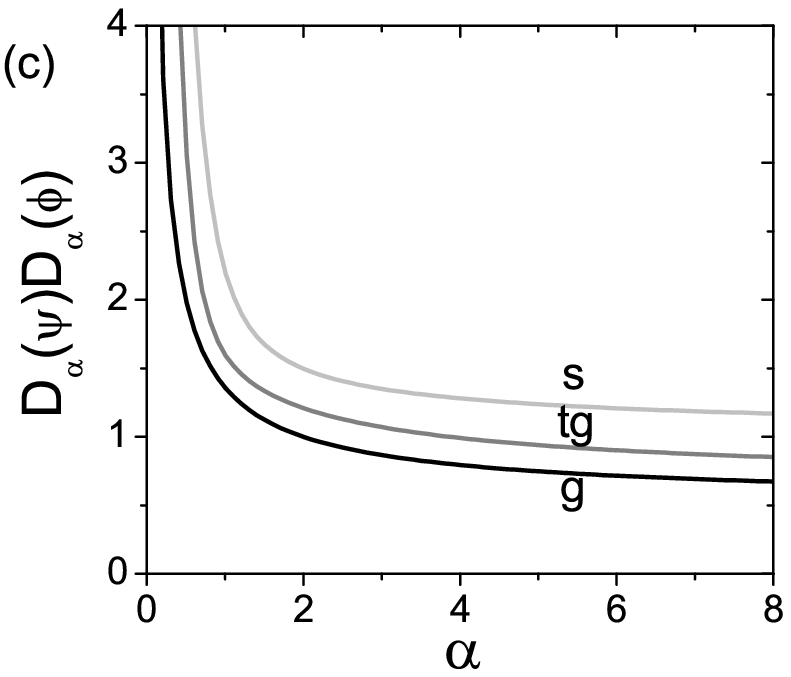}
\caption{Entropic beam products $D_\alpha(\psi)D_\alpha(\phi)$ of light beams with near-field (a) Gaussian profile $\psi(x)\propto\exp(-x^2)$ [g], super-Gaussian profile $\psi(x)\propto\exp(-x^8)$ [sg], slit profile $\psi(x)\propto 1$ in $[-1,1]$ [s], with (b) Hermite-Gauss profiles $\psi(x)\propto\exp(-x^2)H_m(\sqrt{2}x)$ of the indicated orders $m=0,1,2$ and $3$, and with (c) the truncated Gaussian profile $\psi(x)\propto\exp(-x^2)$ in $[-1,1]$ [tg] compared to those with Gaussian and slit profiles.}
\label{Fig3}
\end{figure*}

In the light of the above, inequality (\ref{ENTRO3}) for two light beams such that $r\succ \tilde r$, can be rewritten, upon exponentiation, as
\begin{equation}\label{ENTROPRO}
D_\alpha(\psi)D_\alpha(\phi)\le D_\alpha(\tilde\psi) D_\alpha(\tilde \phi)\,.
\end{equation}
Majorization then implies that the product of all entropic widths in space and spatial frequency, or entropic beam width products, are smaller for the majorizing beam than for the majorized beam. In practice, the products $D_\alpha(\psi)D_\alpha(\phi)$ can directly be calculated from the discretized data as $\Delta x\Delta\xi \exp[H_\alpha(r)]$, or, equivalently $\Delta x\Delta\xi \left(\sum_{j=1}^{NN'} r^\alpha\right)^{1/(1-\alpha)}$. Relation (\ref{ENTROPRO}) supports the affirmation that a beam whose Lorenz curve is higher than the Lorenz curve of a second beam can be considered of better quality in a strong sense. As a few examples, Fig. \ref{Fig3} shows the product of the entropic widths for the beams whose Lorenz curves in Fig. \ref{Fig1} keep the majorization relation.

\section{Intersecting Lorenz curves and contradicting entropic beam width products}

Naturally, a majorization relation between two beams may not hold, i. e., their Lorenz curves cross each other, as considered below. Lack of majorization relation between two beams just tell us that one must resort to particular criteria, selected according to the particular application, to evaluate whether a beam can be considered of better quality than a second beam or vice versa. The effects of lack of majorization have been studied in other contexts in \cite{LUIS2}.

A related question is whether exists a beam that majorizes any other beam, that is, an optimal quality beam. Recent developments \cite{PUCHALA,FRIEDLAND} have shown that there exists a vector $w$ such that $w\succ r=q\otimes p$ for any $p$ and $q$ extracted from $\psi(x)$ and $\phi(\xi)$. The existence of $w$ implies, in particular, that $H_\alpha(r)\ge H_\alpha(w)$ for any $r$ (and the same for any Schur-concave function), and therefore $D_\alpha(\psi) D_\alpha(\phi)\ge B_\alpha$, where $B_\alpha=\Delta x\Delta \xi \exp[H_\alpha(w)]$ are lower bounds, whose values are the subject of an intense debate \cite{DEUTSCH,WEHNER,PARTOVI,PUCHALA,FRIEDLAND,RASTEGINI,RUDNICKI}. However, the evaluation of $w$ is complicated even in low dimensional cases (few points $NN'$). Further, in our context we do not consider any all "probability" distributions [such as e. g., $q=(1/N,\dots, ,1/N)$, $p=(1,0,\dots, 0)$, that does not represent the discretization of a beam whose total power fall into the detectors], but only distributions simulating continuous and localized functions in the unbounded space $(x,\xi)$, and these subset of distributions may not saturate those lower bounds.

\begin{figure}[b!]
\centering
\includegraphics[width=4.5cm]{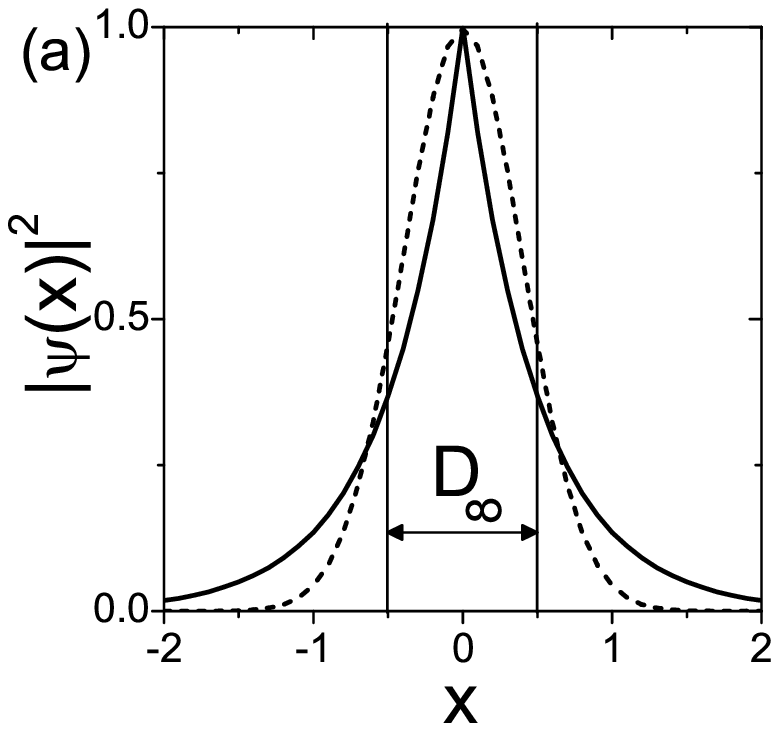}\includegraphics[width=4.5cm]{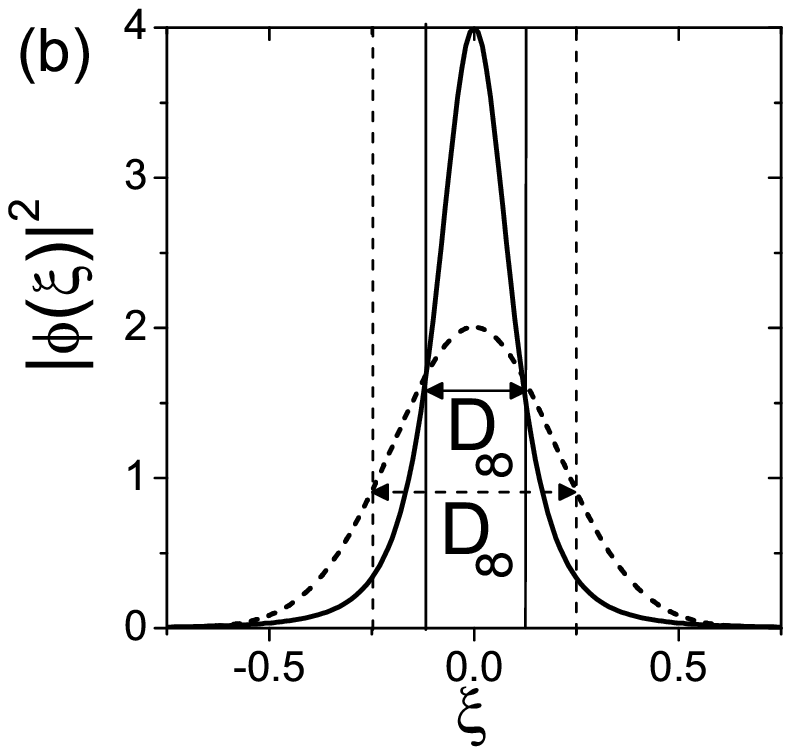}
\includegraphics[width=4.5cm]{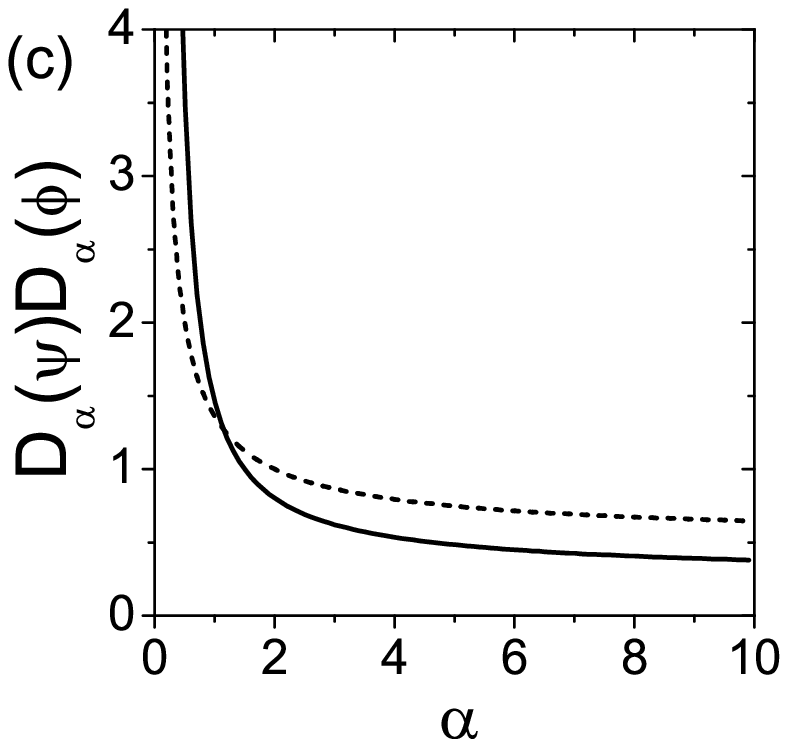}\includegraphics[width=4.5cm]{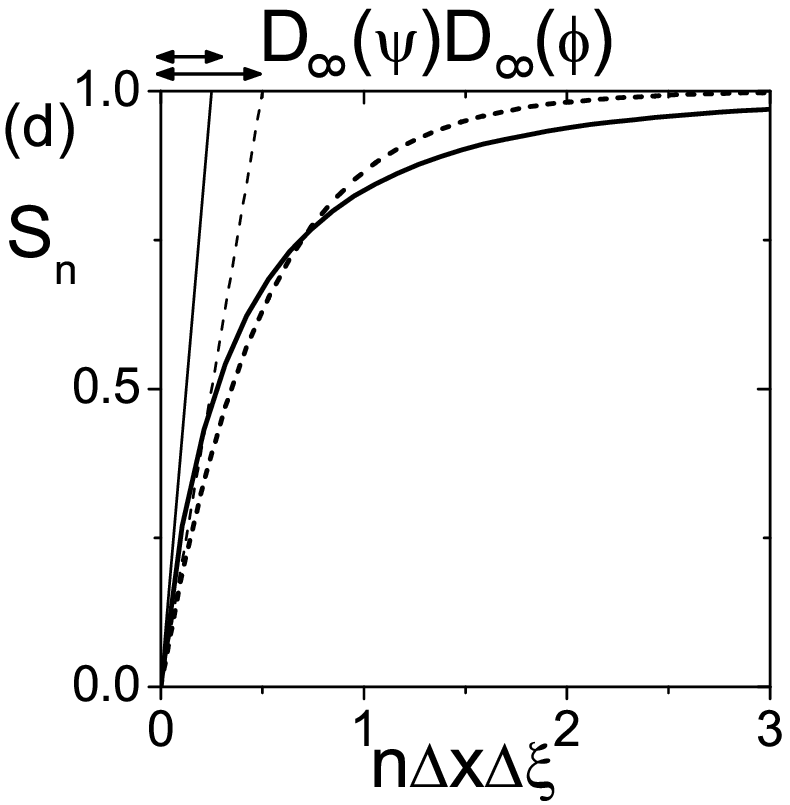}
\caption{(a) Intensity profiles of the Gaussian beam $\psi_1(x)=(2/\pi a_1^2)^{1/4}\exp(-x^2/a_1^2)$, $a_1=0.8$ (dashed curve), and of the two-side exponential beam $\psi_2(x)=\exp(-|x|/a_2)/a_2^{1/2}$, $a_2=1$ (solid curve) with unit power, $P=1$, and unit peak intensity, and therefore with the same min-entropic width $D_\infty=1$. (b) Intensity profiles of their Fourier transforms $\phi_1(\xi)=(2\pi a_1^2)^{1/4}\exp[-(\pi a_1\xi)^2]$ (dashed curve), and $\phi_2(\xi)=2a_2^{1/2}/[1+(2\pi a_2\xi)^2]$ (solid curve), also with unit power, $P=1$, but different peak internsities, $2$ and $4$, respectively, and therefore, min-entropic widths $0.5$ and $0.25$, also respectively. (c) Entropic beam products and (d) Lorenz curves for the Gaussian beam (dashed curves) and for the two-side exponential beam (solid curves). The straight lines in (d) are $n\Delta x\Delta \xi/D_\infty(\psi)D_\infty(\phi)$, reaching unit values at the corresponding min-entropic width products $D_\infty(\psi)D_\infty(\phi)$.}
\label{Fig4}
\end{figure}

For such distributions, Gaussian $q$ and $p$ distributions correspond to a Gaussian beam, which is usually considered the best light beam. This assertion is true from the perspective of variance-based widths \cite{SIEGMAN1,WRIGHT}, or from Shannon entropies \cite{BIRULA1}, but not from other perspectives, as is clear from the following example. In Fig. \ref{Fig4}(a) the Gaussian profile $\psi_1(x)\propto \exp(-x^2/a_1^2)$ and the double-side exponential profile $\psi_2(x)\propto \exp(-|x|/a_2)$ are scaled vertically and horizontally so that they have the same power and peak intensity, and therefore also have the same min-entropic widths, $D_\infty(\psi_1)=D_\infty(\psi_2)$. The corresponding far-field distributions, i. e., the Gaussian profile $\phi_1(\xi)\propto\exp[-(\pi a_1\xi)^2]$ and the Lorenzian profile $\phi_2(\xi)\propto 1/[1+(2\pi a_2\xi)^2]$, are plotted in Fig. \ref{Fig4}(b). They also have the same power, but the peak intensity of the Gaussian is one half that of the Lorenzian, and therefore the min-entropic width of the far-field Gaussian profile is twice that of the Lorenzian profile, $D_\infty(\phi_1)=2D_\infty(\phi_2)$. It is not hard to admit, looking at Fig. \ref{Fig4} (a) that the Gaussian and the two-side exponential profiles have indeed similar widths, and looking at Fig. \ref{Fig4} (b), that the Lorenzian profile is narrower than the Gaussian profile, the entropic widths thus providing quite good estimates of their widths both in space and spatial frequency (close to the standard full-widths at half maxima). Consequently, the min-entropic width product $D_\infty(\psi)D_\infty(\phi)$ of a Gaussian beam is twice that of the two-side exponential beam, $D_\infty(\psi_1)D_\infty(\phi_1)=2 D_\infty(\psi_2)D_\infty(\phi_2)$, the former being, according to this criterion, of lower quality than the latter. If, however, one calculates the Shannon entropic width products, one obtains the opposite conclusion that $D_1(\psi_1)D_1(\phi_1)=0.927 D_1(\psi_2)D_1(\phi_2)$ (for the variance-based product the factor is $0.854$). Figure \ref{Fig4}(c) plots all entropic width products for the Gaussian and for the two-side exponential beams. The existence of contradicting results with different entropic widths implies that there is no a majorization relation between these two beams, and therefore that the Gaussian beam cannot be considered of absolutely better quality than the two-side exponential beam.

All this information can be understood, at least qualitatively, by taking a look at the respective Lorenz curves, which, as expected, intersect [Fig. \ref{Fig4}(d)]. In particular, the value of the min-entropic beam product of a light beam can directly be extracted from the Lorenz curve: Note that the first element of the partial sums is $S_1=r^{\downarrow}_1=\mbox{max} \, r=\mbox{max}\{|\psi(x)|^2|\phi(\xi)|^2\}\Delta x \Delta\xi/P^2 = \Delta x\Delta\xi/D_\infty(\psi)D_\infty(\phi)$. So, the straight line $n\Delta x\Delta \xi/D_\infty(\psi)D_\infty(\phi)$ with the same initial "slope" as the Lorenz curve reaches unity at the min-entropic width product $D_\infty(\psi)D_\infty(\phi)$, as illustrated in Fig. \ref{Fig4}(d). In the example of Fig. \ref{Fig4}, the higher initial slope for the two-side exponential beam informs us that the highest intensities in both the near and far-fields are more concentrated in the two-side exponential beam than in the Gaussian beam, i. e., the min-entropic width product, and other entropic products with high values of $\alpha$, are smaller. Conversely, the lower "tail" of the Lorenz curve of the two-side exponential beam also informs us that the two-side exponential beam has, at the same time, more widespread intensity at low levels than the Gaussian beam, and therefore the entropic products for lower values of $\alpha$ are larger. In situations such as those shown in this example, one beam cannot be considered of absolutely better quality than the other beam, but only according to specific definitions of beam width.

\begin{figure}[t]
\centering
\includegraphics[width=4.5cm]{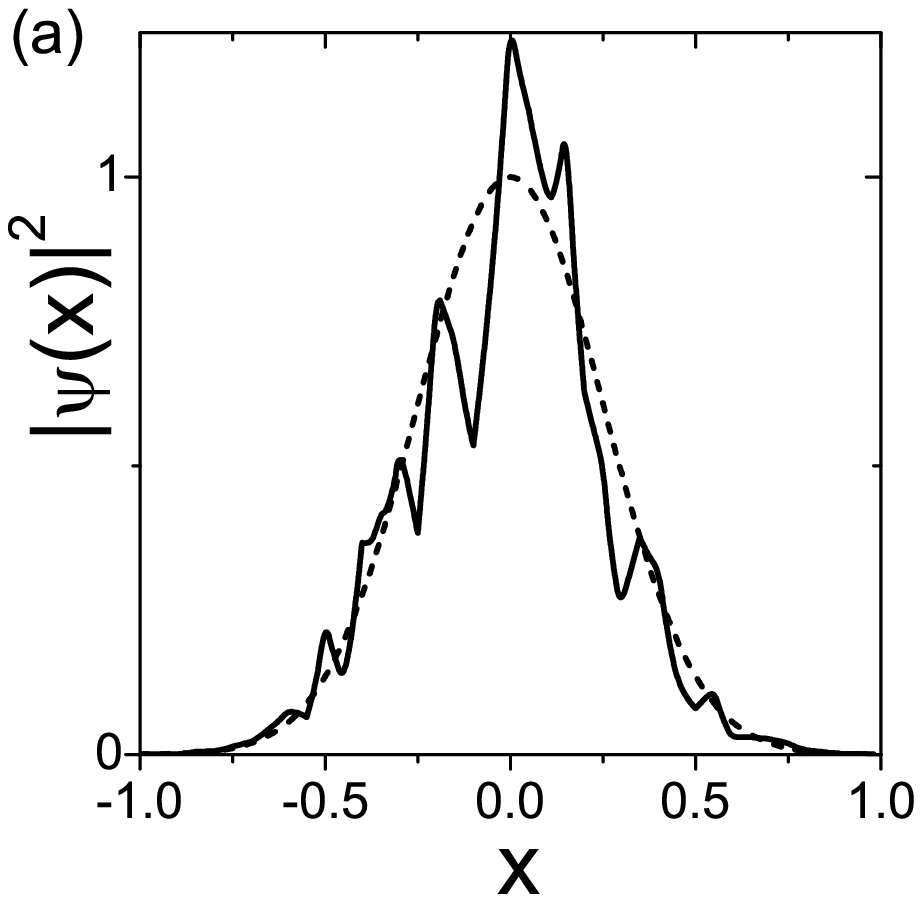}\includegraphics[width=4.5cm]{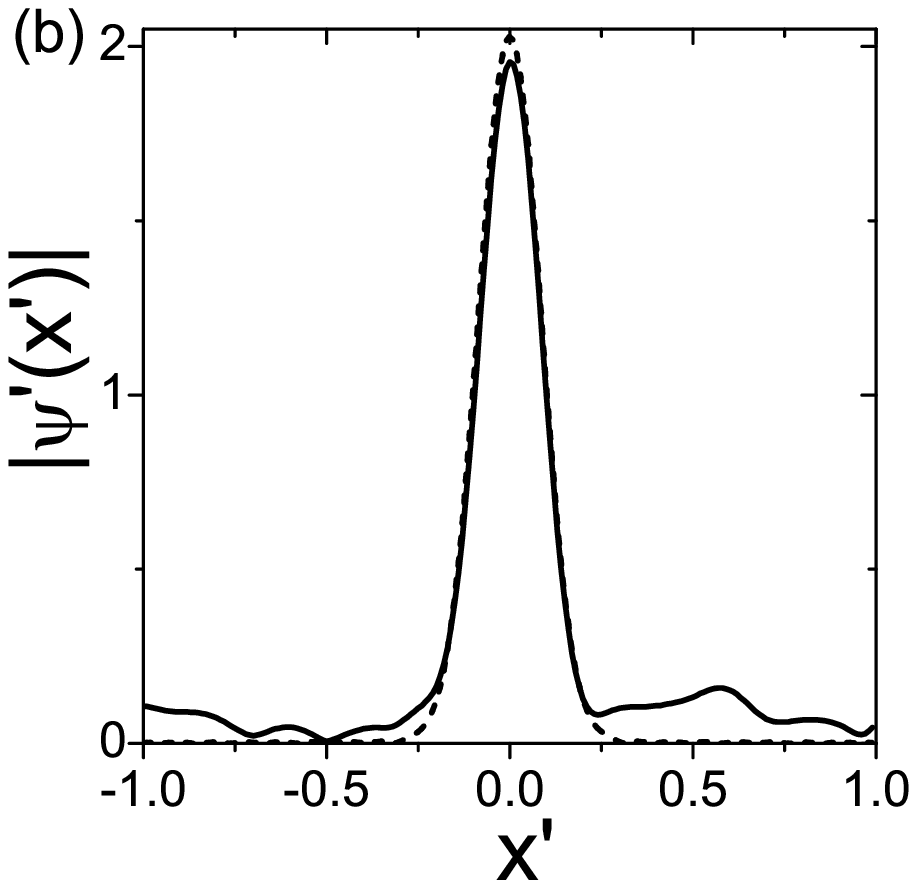}
\includegraphics[height=4.5cm]{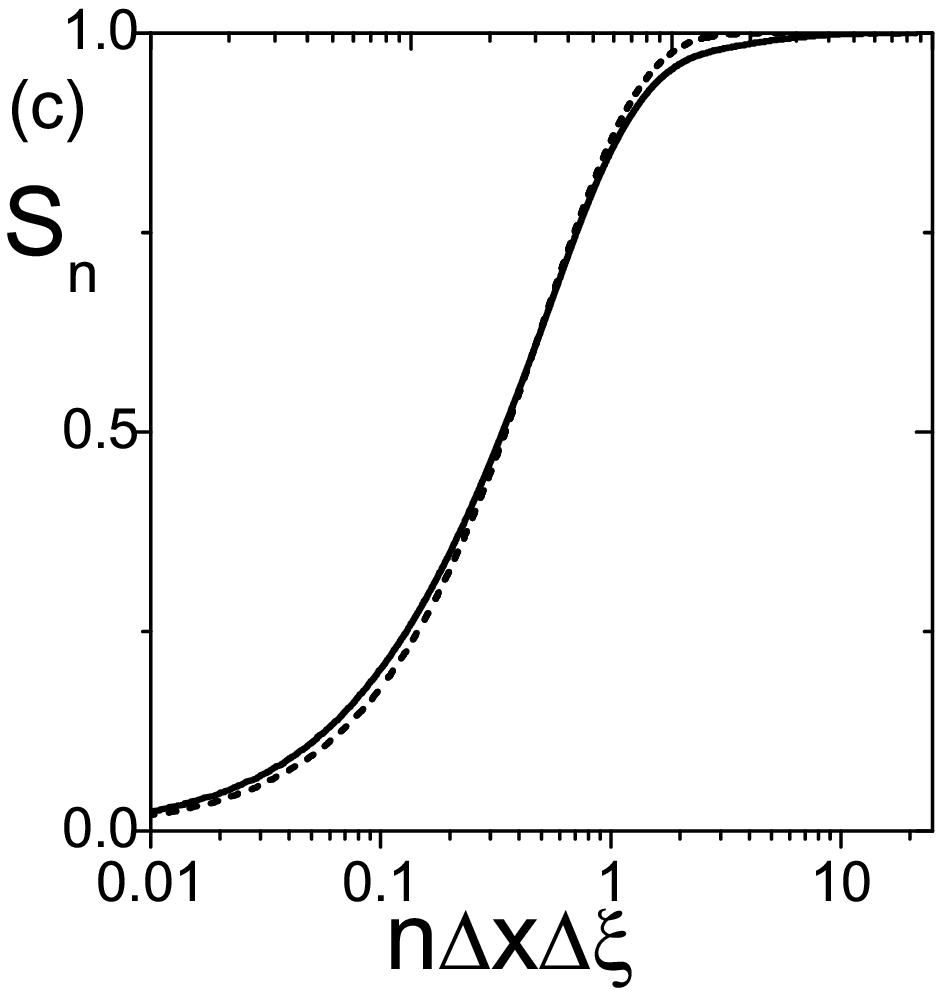}\includegraphics[height=4.5cm]{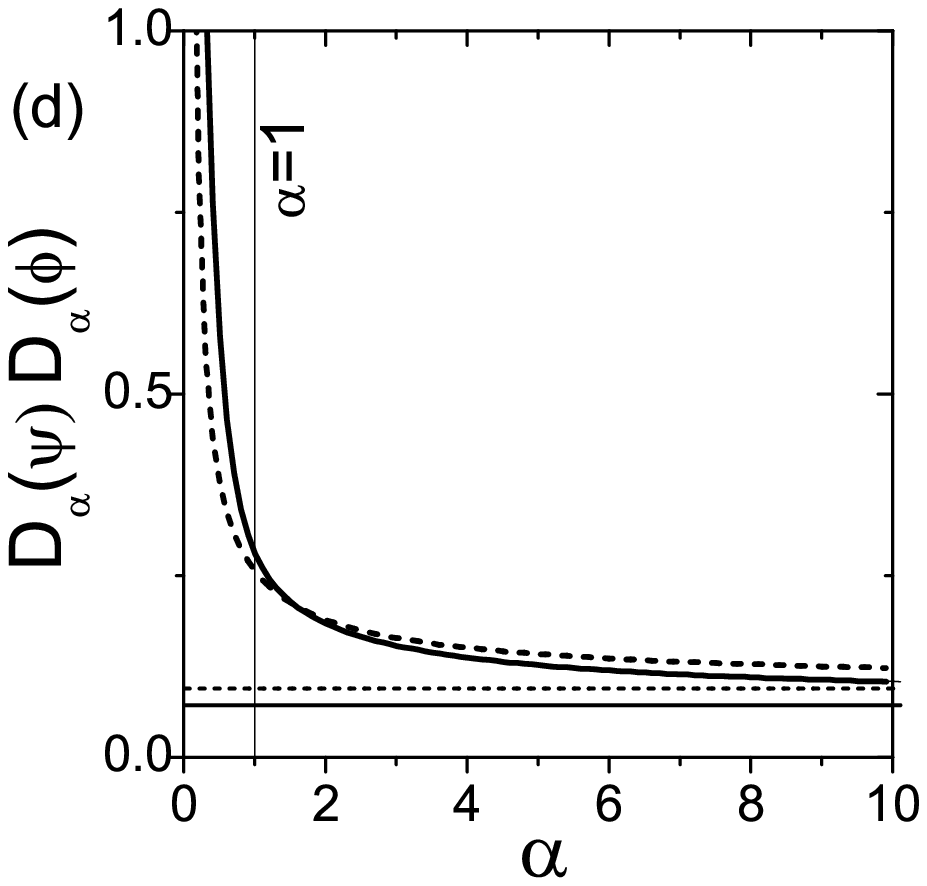}
\caption{(a) Intensity profile of the ideal Gaussian beam $\psi(x)=\exp(-x^2/a^2)$, $a=0.5$ mm and of not-unrealistic Gaussian-like beam (solid curve), obtained by perturbing the amplitude and phase of the Gaussian beam randomly. (b) For light wave length $\lambda=630$ nm,   corresponding amplitude profiles at the focal plane. The amplitude is shown for better visibility of widespread low intensity. With the same pixel size $\Delta x=\Delta x'=10\,\mu$m in front of the lens and at the focal plane (i. e., same CCD camera),  corresponding (c) Lorenz curves and (d) entropic beam products. The logarithmic horizontal scale in (c) enhances the differences between the Lorenz curves. The horizontal lines in (d) are the corresponding min-entropic with products $D_\infty(\psi)D_\infty(\phi)$.}
\label{Fig5}
\end{figure}

Although the Gaussian beam does not majorize all other beams, as seen in the preceding example, the Gaussian beam still plays a prominent role in beam quality considerations. The entropic uncertainty relation for Shannon entropies ($\alpha=1$) introduced in \cite{BIRULA1} states (translated to the optics language) that $H_1(\psi) + H_1(\phi)\ge 1-\ln 2$ for any beam, and that this inequality is only saturated by the Gaussian beam. In terms of the entropic width product, this uncertainty relation reads as $D_1(\psi)D_1(\phi)\ge e/2$, with the equal sign holding only for the Gaussian beam. The existence of a single Schur-concave function ---the Shannon entropy--- that is larger for any beam than for a Gaussian beam implies that no light beam can majorize the Gaussian beam.

We thus propose, in a practical beam quality characterization, to depict the Lorenz curve of the laser beam under consideration together with that of the ideal Gaussian beam for comparison, as in the ``not-unrealistic" example of Fig. \ref{Fig5}. It simulates the output beam from a laser at $\lambda=630$ nm that is focused with a lens of focal length $f=300$ mm. Weak focusing allows proper resolution of the output beam profile and at the focal plane with the same pixel size of $\Delta=10\,\mu$m. The output jagged Gaussian beam and focal distribution in Figs. \ref{Fig5}(a) and (b) (simulated by introducing fluctuations in the amplitude and phase of a Gaussian function) is not as a bad beam as its variance-based quality factor, $M^2=1.71$, suggests. The fluctuations are reflected in a low intensity pedestal at high spatial frequencies, but they do not suffice to lower or broaden significantly the central lobe at the focal plane. All together, the joint intensity distribution has higher and equally concentrated high-intensity values, and at the same time more widespread low intensity-values. These features are reflected in the Lorenz curve in Fig. \ref{Fig5}(c), which has a faster initial rise, but a lower tail, than that of the ideal Gaussian beam. As seen in Fig. \ref{Fig5}(d), lack of majorization results in that entropic width products of large (small) $\alpha$ are smaller (larger) for the jagged Gaussian beam than for the ideal one. All these subtleties can not be captured by a simple parameter like $M^2$, but can be inferred from the Lorenz curve.

\section{Conclusions}

To summarize, we have used recently developed techniques for the quantification of uncertainty to introduce an alternative characterization of the quality of a light beam, in the sense of being more localized in near and far-fields. The most relevant conceptual difference with previous approaches is that the quality is characterized by means of a curve and not by means of parameters. If the Lorenz curve of a light beam is higher than the Lorenz curve of another beam, our approach suggests to consider the former as being of absolutely better quality. This is supported by the fact that all entropic beam width products, measuring the area occupied by the beam in phase space, are smaller for the first beam than for the second beam.
Contrary to the popular parametric characterization of beam quality based on variances, the present method:
\begin{itemize}
\item can be applied to any real beam carrying finite power, including hard-aperture diffracted beams,
\item does not rely on particular beam width definition, and
\item does not introduce any arbitrary criterion.
\end{itemize}
It is just when the Lorenz curves of two beams intersect each other that one beam can be considered of better or lower quality according to different definitions of width or particular criteria, and one must go into these details. In this case, we have also shown how to interpret the intersecting Lorenz curves to ascertain in which sense each of the two beams can be considered of better quality. Given the relevant role of Gaussian beams, we propose to specify the Lorenz curve of the laser beam under consideration together with that of a Gaussian beam for reference.

\bigskip

\noindent\textbf{Funding.} Projects of the Spanish Ministerio de Econom\'{\i}a y Competitividad No. MTM2015-63914-P (M. A. P.) and No. FIS2016-76110-P (I. G.).

\noindent \textbf{Acknowledgment.} The authors thank A. Luis for useful discussions and suggestions.

\section*{Appendix}

We briefly demonstrate that the Lorenz curve of the joint intensity distribution reduces to that defined in Ref. \cite{LUIS} for the focal distribution of uniformly illuminated apertures. In our one-dimensional configuration, suppose $\psi(x)$ is a slit of width $a$, or more generally, a set of slits of total width $a$. The number of illuminated pixels is $N_a=a/\Delta x$, so that $q$ contains $N_a$ times the value $1/N_a$ and $N-N_a$ times the value $0$. We next define $q^{(a)}=(1/N_a,\dots N_a\,\, \mbox{times} \dots, 1/N_a)$. A well-known property of probability distributions is that the Lorenz curve of $q^{(a)}_jp_k$ is the same as that of $p_k$, but the later is sub-sampled by a factor $N_a=a/\Delta x$, as if there were a single pixel in the whole illuminated area in the $x$ plane. At the same time, $q_j p_k$ only differs from $q^{(a)}_jp_k$ in that $q_j p_k$ has additional $(N-N_a)N'$ zero entries. This means that the Lorenz curve of $q_jp_k$ is the same as that of $q^{(a)}_jp_k$ except for additional $N-N_a$ irrelevant unit values forming a horizontal tail. Thus, the Lorenz curve defined by the points $(S_n, n\Delta x\Delta x'/\lambda f)$ is the same as that defined by the sparser points $(S'_n, n a \Delta x/\lambda f)$ except for unit values.

\end{document}